\documentclass[useAMS,usenatbib]{mn2e}

\newcommand{\FIG}[1]{}

\usepackage{amssymb}
\usepackage{amsmath}
\usepackage{subfigure}
\usepackage{textcomp}
\usepackage{graphicx}

\title{Formation of multiple low mass stars, brown dwarfs and planemos via gravitational collapse}

\author[R. Riaz, S. Vanaverbeke and  D.R.G. Schleicher ]{R. Riaz$^{1}$\thanks{E-mail: rriaz@astro-udec.cl} 
S. Vanaverbeke$^{2}$\thanks{E-mail: siegfriedvanaverbeke@gmail.com} D.R.G. Schleicher$^{1}$\thanks{E-mail: dschleicher@astro-udec.cl}\\
$^{1}$Departamento de Astronom\'ia, Facultad Ciencias F\'isicas y Matem\'aticas, Universidad de Concepci\'on, Av. Esteban Iturra s/n Barrio \\
Universitario, Casilla $160$-C, Concepci\'on, Chile \\
$^{2}$Centre for mathematical Plasma-Astrophysics, Department of Mathematics, KU Leuven, Celestijnenlaan 200B, 3001 Heverlee, Belgium \\ 
}

\begin{document}

\date{Accepted - Received -}

\pagerange{\pageref{firstpage}--\pageref{lastpage}} \pubyear{2015}

\maketitle

\label{firstpage}


\begin{abstract}
The origin of very low-mass stars (VLMS) and brown dwarfs (BDs) is still an unresolved topic of star formation. We here present numerical simulations of the formation of VLMS, BDs, and planet mass objects (planemos) resulting from the gravitational collapse and fragmentation of solar mass molecular cores with varying rotation rates and initial density perturbations. Our simulations yield various types of binary systems including the combinations VLMS-VLMS, BD-BD, planemo-planemo, VLMS-BD, VLMS-planemos, BD-planemo. Our scheme successfully addresses the formation of wide VLMS and BD binaries with semi-major axis up to 441 AU and produces a spectrum of mass ratios closer to the observed mass ratio distribution (\textit{q} $>$ 0.5). Molecular cores with moderate values of  the ratio of kinetic to gravitational potential energy (0.16 $\leq$ $\beta{}$ $\leq$ 0.21) produce planemos. Solar mass cores with rotational parameters $\beta{}$ outside of this range yield either VLMS/BDs or a combination of both. With regard to the mass ratios we find that for both types of binary systems the mass ratio distribution varies in the range 0.31 $\leq$ $q$ $\leq$ 0.74. We note that in the presence of radiative feedback, the length scale of fragmentation would increase by approximately two orders of magnitude, implying that the formation of binaries may be efficient for wide orbits, while being suppressed for short-orbit systems.

\end{abstract}

\begin{keywords}
molecular clouds, gravitational collapse, stellar dynamics, low mass binaries, brown dwarfs 
\end{keywords}

\section{Introduction}

Stars and planets are among the most important classes of self-gravitating objects in the Universe. When viewed in terms of their masses, stars and planets occupy distinct regions of the parameter space. Yet nature seems to make a bridge between stars and planets in the form of another class of objects with intermediate mass which are called brown dwarfs (BDs). Brown dwarfs are defined as self-gravitating objects with masses below the minimum mass for hydrogen burning \citep[0.075 $M_{\odot}$ for solar composition models,][]{b10}, yet massive enough to ignite deuterium burning for at least a short episode in their evolutionary history \citep{b9}. Any object whose mass is beyond the hydrogen burning limit is considered a star. Similarly, planetary mass objects or planemos are defined as objects with masses below the deuterium burning limit \citep[$M\leq0.015\,M_\odot$, see][]{b40}
and are assumed to be formed either by accretion of solids or gravitational instabilities in disks orbiting stars or brown dwarfs \citep{b6}. 
In this paper we define very low mass stars (VLMS) as stars with M $\leq{}$ 0.15 $M_{\odot}$ as suggested by \citet{b25}. 
At the upper end of the mass range for brown dwarfs, many objects have been found \citep{b16}. This observation suggests that there exists a continuous transition between brown dwarfs and stars \citep{b16}. On the other hand, observations of the coldest brown dwarfs and Jupiter mass exoplanets indicate 
a temperature gap of several~hundred Kelvin between the two classes of objects \citep{b17}. Since studies of brown dwarf systems provide a link between the theories of star and planet formation, various formation schemes have
been suggested so far to explain how brown dwarfs are formed \citep{b51,b37}.
 
Among the various hypotheses, the embryo-ejection hypothesis has been proposed as a possible scheme to explain the existence of brown dwarfs. It is based on the concept of competitive accretion during the process of star formation. Unlike the traditional picture of an isolated protostar forming within a molecular cloud core, the ejection model assumes that small groups of up to 4 objects are formed inside the collapsing core \citep{b35,b5}. Every single evolving protostar then competes with the others to accrete as much material as possible from the parent envelope of gas. As a result, a fragment which is more massive initially grows more rapidly than the rest. Since the separation between fragments inside the core is small, dynamical interaction comes into play and results in the ejection of the lowest mass object in the group \citep{b41,b47}. If such an ejected fragment, which is no longer able to accrete from its parent cloud, has a mass below the limit of hydrogen burning, it will very likely become a brown dwarf. According to this ejection model, the BD binary fraction is expected to be less than 5$\%$ \citep{b5}. On the other hand, BD binary systems with very low masses are going to be preferentially destroyed during the ejection process. 

The predictions of this model, however, are not generally consistent with observations and with the numerical simulations we will present in this paper.
Observational studies have revealed strong evidence that the formation processes of brown dwarfs 
and low mass stars are in fact very similar. These observations include the spatial
distribution of stars and brown dwarfs in star-forming regions \citep{b22}, the presence of accretion and the frequency of disks orbiting brown dwarfs \citep{b27}, and most importantly for the present study, 
evidence for an embedded phase within molecular cloud cores during brown dwarf formation \citep{b53}. 
The idea of a gravitational collapse scenario for brown dwarf formation is strengthened by many observations of activity surrounding brown dwarfs which is normally related to
gravitational cloud collapse and star formation. The discovery of optical jets around evolving brown
dwarfs \citep{b50} as well as the presence of brown dwarf disks \citep{b21} provides strong clues that a reservoir of material is available around these objects which allows for accretion onto brown dwarfs.

In addition, observations with sub-millimeter arrays have found evidence that brown dwarfs undergo a phase
of bipolar molecular outflow which is typically associated with protostars \citep{b31,b30}.
Similar observational evidence in many wavelength domains has shown
both T Tauri-like accretion \citep{b29,b39,b38} and mass ejection features suggesting that young brown dwarfs are very much behaving like low mass protostars. The presence of jets may also explain why it is difficult for brown dwarfs to reach masses close to the hydrogen burning limit \citep{b2,b49}. 
All of this evidence points toward gravitationally collapsing molecular cores as the birthplace of brown dwarfs and brown dwarf binaries, even if they are formed in isolated parts of the clouds. An example is the binary brown dwarf system {FU Tau\/} which, although regarded as a member of the Taurus molecular cloud, has been found to be situated in a relatively isolated region of the cloud \citep{b28}.
For this object, the spectral types of the components have also been determined
and reveal that {FU Tau\/} is an unequal mass binary system with masses 0.05 $M_{\odot}$ and
0.015 $M_{\odot}$ for {FU Tau A \& B}, respectively \citep{b20}.

Hubble Space Telescope observations and ground-based adaptive optics imaging have
already resolved a considerable number of brown dwarf binary systems and according to a recent estimate,
about 17-30$\%$ of the field brown dwarfs are in fact members of close binary systems \citep{b14}. 
On the other hand, unequal mass brown dwarf binary systems have been found to be rare \citep{b18}. Their existence, however, certainly requires attention because it could be critical to understand the formation mechanism(s) and the physical processes which govern the evolution of these systems. 
The discovery of objects like {2MASS J12095613-1004008\/} has drawn attention towards
these unequal mass binary systems. The low mass ratio (q $\approx{}$0.5) in this particular system has been inferred from the large IR brightness difference between the two components \citep{b11}. On the other hand, tentative evidence has been found for a companion orbiting the L4 brown dwarf {2MASSW 033703-175807\/} with an uncommonly low mass ratio (q $\approx{}$0.2) and a very low temperature \citep[\textit{T}$_{eff}$ $\approx{}$ 600--630 K,][]{b42}, whose existence has so far defied theoretical explanation. 
A last example is the discovery of the binary brown dwarf {WISE 1049-5319\/} with the Wide-field Infrared Survey Explorer (WISE) \citep{b54}, which hosts a low mass secondary
that could be a late L or early T spectral type brown dwarf. Remarkably, this system is the third closest neighbour of the Sun which has been found so far \citep{b19}.

Also, there has been a considerable effort to understand numerically the formation mechanism of brown dwarf both by the scenario of disk fragmentation \citep{Kraus11, Stamatellos08, Stamatellos07, Goodwin07, b5} and as ejected stellar embryos \citep{bate2012, basu12, b45, Reipurth01}. 

All of these observations and numerical findings set the stage for the study that we will present in this paper.
We have attempted to model the initial
stages of the formation of unequal mass brown dwarf and very low mass star systems evolving within a common envelope of gas. 
Our models start with molecular cloud cores with various rotation rates and are seeded with non-axisymmetric density perturbations to mimic the presence of large scale turbulence on the level of molecular cloud cores. We also address the formation and early stages of evolution of binary systems.
We note that our simulations do not incorporate radiative feedback, and thus only accurately describe the dynamics before protostars are formed. After the formation of protostars, the effective equation of state stiffens \citep[see e.g.][]{Offner2009}, suppressing the formation of additional objects. Effectively, the effect of radiative feedback will favor the formation of long-period binaries, while short-period ones will be significantly suppressed. The implications of this are discussed in more detail in section 6.

The plan of the paper is as follows. Section 2 discusses the computational scheme. The initial conditions, the setup for our models and the scheme which is used to identify fragments in the gas and determine the associated mass and other properties are described in section 3. Section 4 discusses the adopted equation of state. Section 5 provides a detailed account of our results and limitations are discussed in section 6. Finally we present our conclusions in section 7.

\section{Computational scheme} \label{scheme}

The numerical models presented in this paper are based on the particle simulation
method known as smoothed particle hydrodynamics (SPH). We use the computer code
GRADSPH \footnote{Webpage GRADSPH: http://www.swmath.org/software/1046} \citep{b46}. GRADSPH is a fully three-dimensional SPH code which combines 
hydrodynamics with self-gravity and has been especially designed to study 
self-gravitating astrophysical systems such as
molecular clouds \citep{b46}.  
The numerical scheme implemented in GRADSPH treats the long range gravitational interactions within the fluid by using a tree-code gravity (TCG) scheme combined with the variable gravitational softening length method \citep{b33}, whereas the short range hydrodynamical interactions are solved using a variable smoothing length formalism.  The code uses artificial viscosity to treat shock waves. 

In SPH, the density $\rho_{i}$ at the position $\vec{r}_{i}$ of each particle with mass $m_{i}$ is determined by summing the contributions from its neighbors using a weighting function 
$W\left(\vec{r}_{i}-\vec{r}_{j},h_{i}\right)$ with smoothing length $h_{i}$:

\begin{equation} \label{density}
\rho _{i} =\sum _{j} m_{j} W\left(\vec{r}_{i}-\vec{r}_{j},h_{i}\right).
\end{equation}

GRADSPH uses the standard cubic spline kernel with compact support within a smoothing sphere of size $2 h_{i}$ \citep{b46,b33}.
The smoothing length $h_{i}$ is determined using the following 
relation: 

\begin{equation} \label{smoothinglength}
h_{i}=\eta \left(\frac{m_{i}}{\rho_{i}}\right)^{1/3}, \ \ \ \ \ 
\end{equation}

where \textit{$\eta$} is a dimensionless parameter which determines 
the size of the smoothing length of the SPH particle given its mass and density. 
The parameter \textit{$\eta$} is derived by requiring that a fixed mass, or equivalently a fixed number of neighbors for equal particle masses,  must be contained inside the smoothing sphere of each particle:

\begin{equation} \label{fixedmass}
\frac{4\pi}{3}\left(2 h_{i}\right)^{3} \rho_{i}=m_{i} N_{opt}=constant. \ \ \ 
\end{equation}

Here N$_{opt}$ denotes the number of neighbors inside the smoothing sphere, which we set equal to
50 for the 3D simulations reported in this paper. 
Substituting Eq. \ref{smoothinglength} into Eq. \ref{fixedmass} allows us to determine $\eta$:

\begin{equation} \label{eta}
\eta=\left(\frac{3 N_{opt}}{32 \pi}\right)^{1/3}.
\end{equation}

Since Eq. \ref{density} depends on $h_{i}$ as defined in Eq. \ref{smoothinglength} and Eq. \ref{smoothinglength} depends on the density defined by Eq. \ref{density}, we need to solve both equations simultaneously using an iterative procedure at each time step and for each particle to determine the smoothing length and the density of the particles given their list of nearest neighbors. The iteration usually converges after only a few steps. A similar procedure is discussed in section 2.4 of the review paper by \citet{Pricereview}.

The evolution of the system of particles is computed using the second-order predict-evaluate-correct (PEC) scheme implemented in GRADSPH, which integrates the SPH form of the equations of hydrodynamics with individual time steps for each particle. For more details and a derivation of the system of SPH equations we refer to \citet{b46}.

\section{Initial conditions and setup} \label{ICs}

In the present study we use a modified version of the Boss and Bodenheimer collapse test with initial conditions described in \citet{b7}. The overall setup of the initial conditions is identical to the cloud core models that we used previously while investigating the thermal response of collapsing molecular cores in \citet{b36} where the total mass inside the core is 1 $M_{\odot}$ and the radius of the core is 0.0162 pc. For all models considered in the present work the gas is initially isothermal with a temperature of $T = 8$~K. The corresponding sound speed is $c_{s}$ = 0.167 km s$^{-1}$. The initial gas density is $\rho_{i} = 3.8 \times 10^{-18}$~g cm$^{-3}$.  The mean free-fall time of the initial condition is given as

\begin{equation} \label{freefalltime}
t_{ff}=\sqrt{\frac{3 \pi}{32 G \rho_{0}}}
\end{equation}

and is 33.968 kyr for the standard initial condition defined above. 

The initial condition is also characterized by the parameters $\alpha$ and $\beta$, which correspond to the ratio of thermal and kinetic energy with respect to the gravitational potential energy of the cloud. These parameters are defined as

\begin{equation} \label{alpha}
\alpha=\frac{5 R k T}{2 G M \mu m_{h}},
\end{equation}

\begin{equation} \label{beta}
\beta=\frac{R^{3}\omega^{2}}{3 G M},
\end{equation} 

where $G$ is the gravitational constant, $k$ is the Boltzmann constant, and $m_{h}$ denotes the mass of the hydrogen atom. In our models, for the initial temperature of 8 K, the initial value of $\alpha$ is kept fixed as 0.265. The initial setup is implemented in our SPH code by placing equal-mass particles on a hexagonal closely packed lattice and retaining only the particles within the initial cloud radius. The code uses internal dimensionless units which are defined by setting G=M=R=1.

We explore values of the rotational parameter $\beta{}$ (the ratio of kinetic energy to gravitational potential energy) within the range 0.0045 $\leq$ $\beta{}$ $\leq$ 0.3267. For the mode of perturbation we use m = 2. Instead of a fixed amplitude A, however, we introduce two distinct amplitudes of the azimuthal density perturbation ($A_{1}$ \& $A_{2}$) in each hemisphere of the uniform density cloud core models. These perturbations are designed to approximately include the effect of large scale turbulence on the scale of giant molecular clouds, which can induce asymmetric density distributions in embedded molecular cores which are on the verge of gravitational collapse. 


\begin{equation} \label{asymperturb1} 
\rho=\rho_{0}\left\{1+A_{1}\sin\left(m\varphi\right)\right\},
\end{equation}

where $\varphi$ is the azimuthal angle in spherical coordinates (r,
$\varphi$, z), for $0 < \varphi < \pi$, and 

\begin{equation} \label{asymperturb2}
\rho=\rho_{0}\left\{1+A_{2}\sin\left(m\varphi\right)\right\},
\end{equation}

for $\pi < \varphi < 2 \pi$, respectively.

An overview of the models is provided in Table 1. The table contains the properties of the initial conditions and relevant information on the final outcome of their evolution. 
By adopting various values for the azimuthal density perturbation amplitudes $A_{1}$ \& $A_{2}$ while keeping the ratios A$_{2}$/A$_{1}$ identical except
for models M2 \& M9, we aim to investigate the formation of asymmetric VLMS and BD binary systems
with possible companions. We do not impose any specific criterion for terminating the models which consider variations in the amplitudes of the initial azimuthal density perturbations. We try to evolve these models as long as possible given the constraint of decreasing timesteps that is caused mainly by the growing gas density.

We also consider a more expensive sub-set of models (M4, M10, M12, M13, M14, \& M15) in which we keep the amplitudes of the initial density perturbation $A_{1}$ \& $A_{2}$ fixed at 0.05 \& 0.025 and vary the rotational parameter $\beta{}$ within a range of 6 values (0.1628, 0.2658, 0.2118, 0.0849, 0.0321, and 0.0045). 
For the set of 6 models defined in this way, we compute the mass which remains in the envelope of the core and has not been accreted by the protostars. The fraction of the total mass of the core which remains in the envelope is used as the criterion for terminating the simulations. We terminate the simulations when the total mass inside the envelope has declined to approximately 70 \% of the initial mass of the core (see Tables 3 \& 4). 
The above two strategies give us the opportunity to investigate numerically the number of secondary fragments (other than those which form directly from the initially introduced density perturbations in the gas) in the embedded phase of core collapse, the mass of the fragments and hence their type (i.e. VLMS, BDs, and planemos) as well as the number of binary systems and their properties, including the binary separation $d$, the semi-major axis $a$, the eccentricity $e$, and the mass ratio $q$. 

We use the following algorithm to identify the fragment(s) formed during the cloud collapse and their associated properties. There are two crucial parameters. The first one is the threshold density $\rho_{threshold}$ above which the algorithm tests for a potential self gravitating fragment. The second one is the parameter $r_{search}$ which determines a search radius around the particle, and looks for all particles within that radius that have densities larger than \textit{$f$} times the density of the particle of interest, where \textit{$f$} = 0.001, and then computes the mass, the energies and the center of mass coordinates and velocities of the fragment on the basis of the particle properties within the search radius. In this work we set $\rho_{threshold} = 10^{-13}$~g cm$^{-3}$ and $r_{search} = 50$~AU.

\section{Equation of state} \label{EOs}

Another modification to our previous models concerns the chemical
composition of the cloud, which is assumed to be a mixture of hydrogen and helium gas
with mean molecular weight $\mu$ = 2.35. We adopt a barotropic equation of state of the form 

\begin{equation} \label{EOS}
P=\rho c_{0}^{2}\left[1+\left(\frac{\rho}{\rho_{crit}}\right)^{\gamma-1}\right],
\end{equation}

with $\gamma$ = 5/3. This approximate equation of state describes the gradual transition from isothermal to adiabatic behavior of the gas during gravitational collapse. For the critical density \textit{$\rho$}$_{crit}$ we use the value 10$^{-13}$ g $cm^{-3}$, which is slightly higher than the value 10$^{-14}$ g $cm^{-3}$ which was used as the fiducial value before in \citet{b36}. Our current choice for the value of \textit{$\rho$}$_{crit}$ = 10$^{-13}$ g $cm^{-3}$ is consistent with \citet{bonnell1994, Larson1969} and \citet{Hayashi1966} who have demonstrated that beyond a density of 10$^{-13}$ g $cm^{-3}$ the gas becomes optically thick and the pressure forces increase faster with density compared to the gravitational forces.

\citet{bateburkert1997} provide the following expression of the maximum resolvable density for an SPH calculation

\begin{equation} \label{Resolvable density}
\rho_{crit}\backsimeq \left(\frac{3}{4 \pi}\right)   \left(\frac{5R_{g}T}{2G \mu}\right)^{3} \left(\frac{N_{tot}}{2N_{neigh} M_{tot}}\right)^{2}.
\end{equation}

We use a total number of 250025 SPH particles in each model. Setting $\rho_{crit}=\rho_{threshold}$ in the above equation leads to a minimum of 119523 SPH particles to meet the resolution criterion. In the present set of simulations the number of SPH particles is more than twice the value which is required to resolve the clumps forming in our simulations hence ensuring numerical consistency and preventing any non-physical fragmentation. 

The approach described above allows us to follow the formation of a protostellar core by making the evolution adiabatic at very high densities. This prevents gravitational collapse to even smaller scales, and therefore allows us to investigate fragmentation and binary formation within the system. It also has the advantage that hydrodynamical interactions are fully taken into account. 

An alternative would be to use a sink particle algorithm as implemented for example by \citet{b12}, which is computationally more efficient since it avoids continually decreasing timesteps \citep[see also][]{Riaz17}, but which intrinsically cannot account for the hydrodynamical interactions of the protostars with the surrounding gas. The approach adopted here will therefore give us a good impression about the dynamical evolution during the earlier stages, but of course is naturally limited by smaller timesteps during the formation of the protostellar cores.

\begin{figure*} \label{fig:1}
    \centering
    \includegraphics[angle=0,scale=0.45]{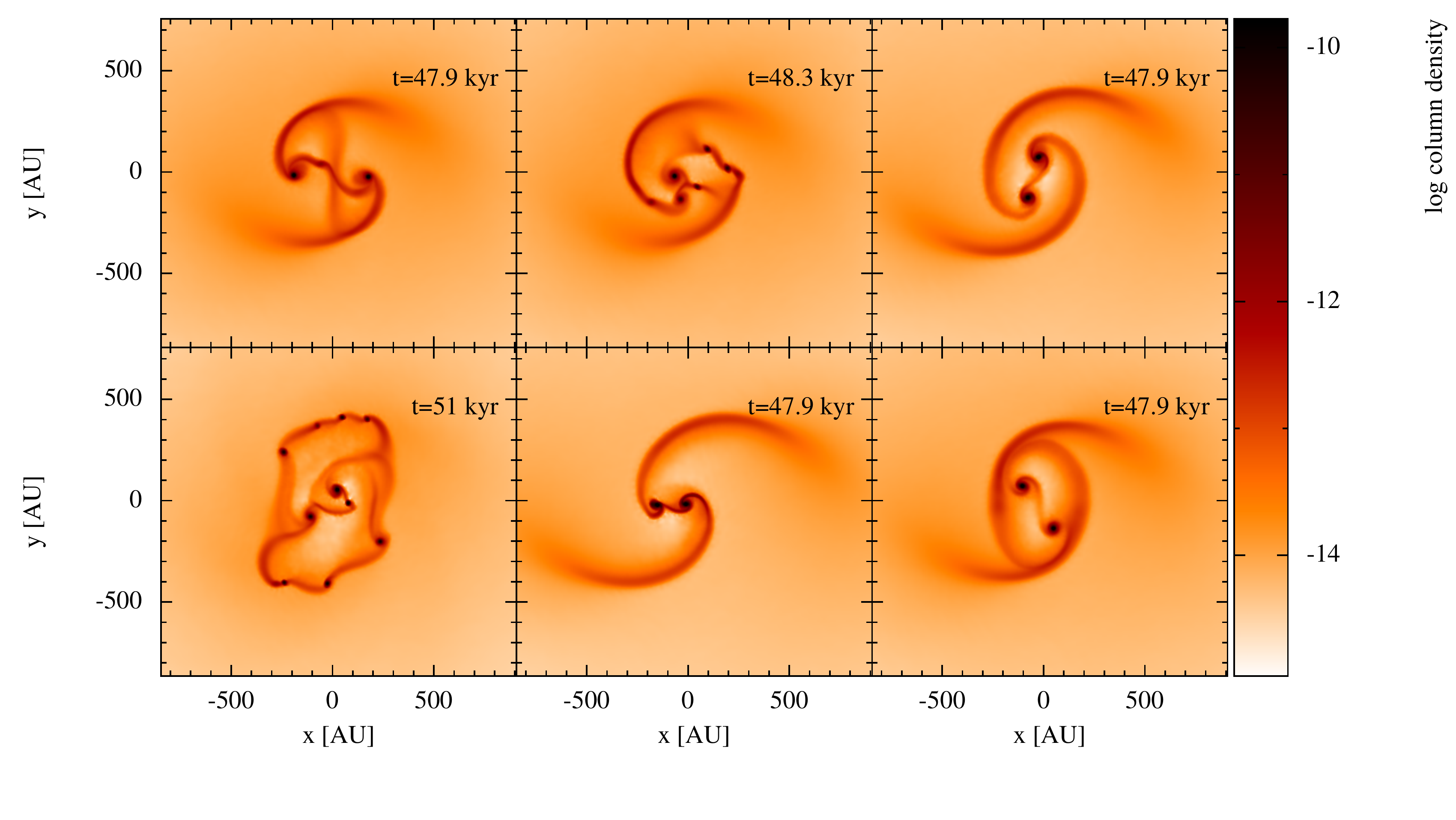}
    \caption{Simulation results for models M1 - M6 (panels from top left to bottom right) at the end of the computed evolution of each model. Each plot shows a face-on view of the column density integrated along the z-axis. The colour bar on the right shows log ($\Sigma{}$) in physical units of g cm$^{-2}$. Each calculation was performed with 250025 SPH particles.}
  \end{figure*}

   \begin{figure*} \label{fig:2}
    \centering
    \includegraphics[angle=0,scale=0.45]{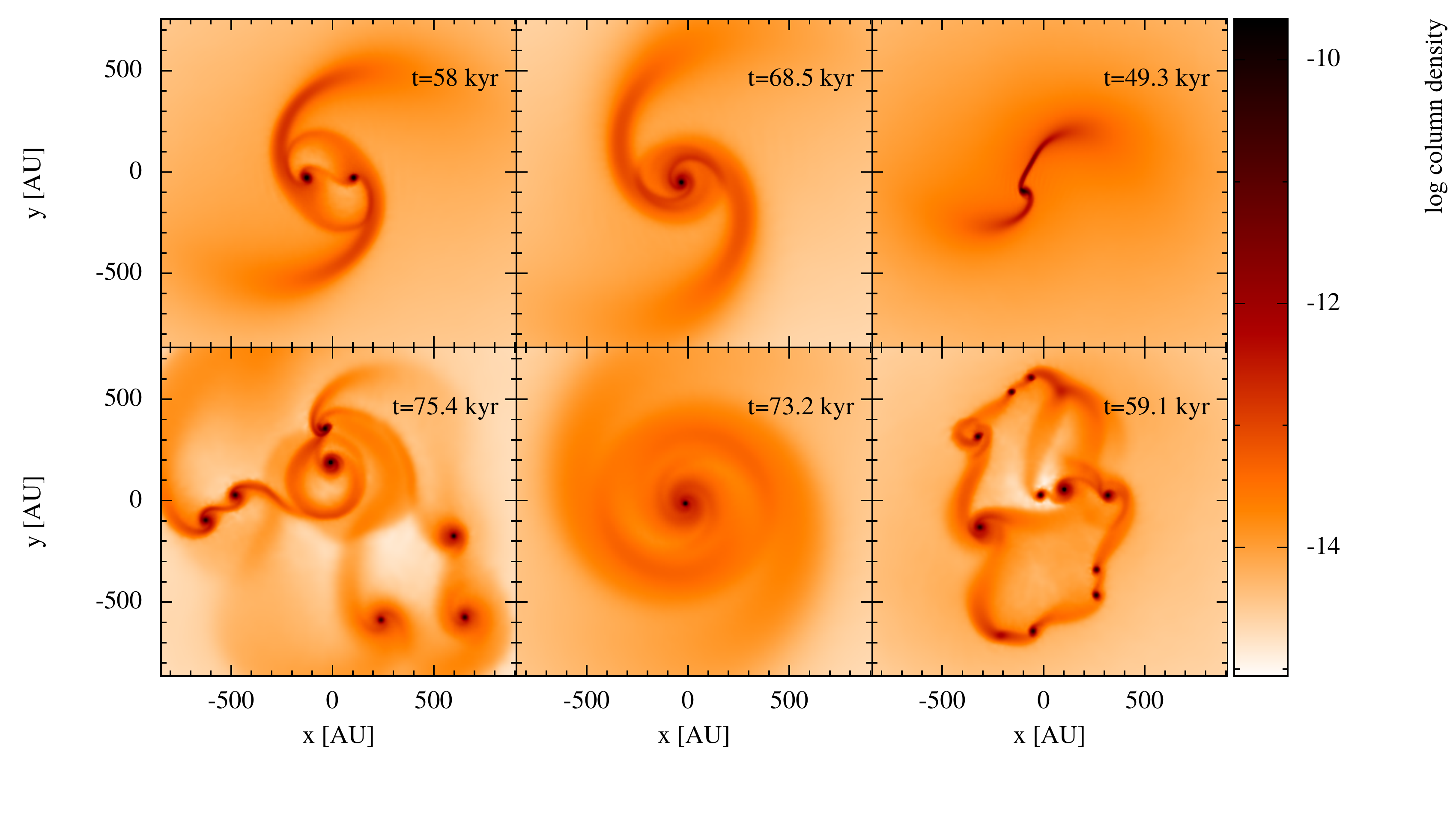}
    \caption{Simulation results for models M7 - M12 (panels from top left to bottom right) at the end of the computed evolution of each model. Each  plot shows a face-on view of the column density integrated along the z-axis. The colour bar on the right shows log ($\Sigma{}$) in physical units of g cm$^{-2}$. Each calculation was performed with 250025 SPH particles.}
  \end{figure*}

\begin{table*} \label{tbl-1}
\centering
\caption{Summary of the initial physical parameters and the final outcome of the models considered in this paper. The initial mass, radius, density and temperature for each model are given by the constant values 5 x 10$^{16}$ cm, 1 $M_{\odot}$, 3.8 x 10$^{-18}$ g $cm^{-3}$ and 8 K, respectively. }
\begin{tabular}{cccccc}
\hline
\hline
Model & ($A_{1}$, $A_{2}$) & Rotational parameter ($\beta{}$) & Rotational speed (rad $s^{-1})$  & Final outcome\\
\hline
M1  & 0.1,0.05   & 0.1628 &  7.2 x 10$^{-13}$  & Single VLMS, single BD, single planemo \\
M2  & 0.1,0.025  & 0.1628 &  7.2 x 10$^{-13}$  & Multiple BDs, two planemos   \\
M3  & 0.2,0.1	    & 0.1628 &  7.2 x 10$^{-13}$  & Two VLMS \\
M4  & 0.05,0.025 & 0.1628 &  7.2 x 10$^{-13}$  & Single VLMS, multiple BDs, single planemo  \\
M5  & 0.25,0.125 & 0.1628 &  7.2 x 10$^{-13}$  & Two VLMS \\
M6  & 0.15,0.075 & 0.1628 &  7.2 x 10$^{-13}$  & Two VLMS \\
M7  & 0.1,0.05   & 0.2658 &  9.2 x 10$^{-13}$  & Two BDs \\
M8  & 0.1,0.05   & 0.3267 &  10.2 x 10$^{-13}$ & Single BD \\
M9  & 0.1,0.025  & 0.2118 &  8.2 x 10$^{-13}$  & Two BDs \\ 
M10 & 0.05,0.025 & 0.2658 &  9.2 x 10$^{-13}$  & Singe VLMS, multiple BDs \\ 
M11 & 0.05,0.025 & 0.3267 &  10.2 x 10$^{-13}$ & Single BD \\ 
M12 & 0.05,0.025 & 0.2118 &  8.2 x 10$^{-13}$  & Single VLMS, multiple BDs, single planemo \\ 
M13 & 0.05,0.025 & 0.0849 &  5.2 x 10$^{-13}$  & Multiple BDs   \\
M14 & 0.05,0.025 & 0.0321 &  3.2 x 10$^{-13}$  & Two VLMS, two BDs   \\
M15 & 0.05,0.025 & 0.0045 &  1.2 x 10$^{-13}$  & Single VLMS   \\
\hline
\end{tabular}
\end{table*}

\begin{table*} \label{tbl-2}
\centering
\caption{Summary of the final evolution time $t_{f}$, the masses of the fragments and the number of VLMS, BDs, and planemos for each model.}
  \begin{tabular}{cccccc}
\hline
\hline
Model & $t_{f}$ (kyr) & Fragment mass ($M_{\odot}$) & VLMS / BDs / planemos\\
\hline
M1  & 47.9  &  0.0116, 0.0773, 0.0674 & 1 / 1 / 1 \\
M2  & 48.3  &  0.0225, 0.0219, 0.0118, 0.0247, 0.0128, 0.0167, 0.0745 & 0 / 5 /2   \\
M3  & 47.9  &  0.1022, 0.0993 & 2 / 0 / 0 \\
M4  & 51.0  &  0.0215, 0.0284, 0.0275, 0.0815, 0.0363, 0.0107, 0.0200, 0.0255, 0.0334, 0.0162   &  1 / 8 / 1  \\
M5  & 47.9  &  0.1098, 0.1085 & 2 / 0 / 0 \\
M6  & 47.9  &  0.0855, 0.0776 & 2 / 0 / 0 \\
M7  & 58.0  &  0.0670, 0.0473 & 0 / 2 / 0 \\
M8  & 68.5  &  0.0634	        & 0 / 1 / 0 \\
M9  & 49.3  &  0.0314, 0.067 & 0 / 2 / 0 \\ 
M10 & 75.4  & 0.0285, 0.0374, 0.0341, 0.0482, 0.0885, 0.0210, 0.0319  & 1 / 6 / 0 \\ 
M11 & 73.2  & 0.0465  &  0 / 1 / 0 \\ 
M12 & 59.1  & 0.0234, 0.0089, 0.0218, 0.0159, 0.0207, 0.0332, 0.0801, 0.0181, 0.0284, 0.0159, 0.0392  & 1 / 9 / 1  \\
M13 & 41.7  & 0.0333, 0.0274, 0.0268, 0.0329, 0.0729, 0.0437, 0.0198, 0.0459  & 0 / 8 / 0 \\ 
M14 & 38.0  & 0.0232, 0.0857, 0.1384, 0.0559  & 2 / 2 / 0 \\ 
M15 & 36.0  & 0.2997  & 1 / 0 / 0  \\ 

\hline
\end{tabular}
\end{table*}

\begin{table*} \label{tbl-3}
\centering
 \caption{Summary of the central density $\rho_{c}$, the central temperature $T_{c}$, the radius of the fragment, the final accretion rate ($\dot M_{\rm final}$) for the most massive fragment, and the ratio of the envelope mass to the initial mass of core for models models M1 - M8.}
 \begin{tabular}{ccccccccc}
\hline
\hline
 Model & M1 & M2 & M3 & M4 & M5 & M6 & M7 & M8 \\ 
\hline 
 $\rho_{c}$(g $cm^{-3}$ x 10$^{-11}$) & 8.33 & 10.34 & 9.22 & 10.19 & 8.70 & 8.77 & 5.56 & 4.65  \\
 
 $T_{c}$(K) & 1088 & 601 & 1486 & 1330 & 1402 & 1146 & 897 & 907  \\
 
 $R_{fragment}$(AU) & 9 & 6 & 10 & 9 & 10 & 9 & 10 & 11  \\
 
 $\dot M_{\rm final}(M_{\odot}$ yr$^{-1}$ x 10$^{-6}$) & 6.698 &  1.812  &  7.680  & 3.041 & 23.680 & 3.527  & 2.224  & 1.822   \\
 $M_{env}/M_{c} (\%)$ & 84.37 & 81.51 & 79.85 & 69.90 & 78.17 & 83.69 & 88.57 & 93.66  \\
\hline
\end{tabular}
\end{table*}

\begin{table*} \label{tbl-4} 
\centering
\caption{Summary of the central density $\rho_{c}$, the central temperature $T_{c}$, the radius of the fragment, the final accretion rate ($\dot M_{\rm final}$) for the most massive fragment, and the ratio of the envelope mass to the initial mass of core for models models M9 - M15.}
\begin{tabular}{ccccccccc}
\hline
\hline
 Model & M9 & M10 & M11 & M12 & M13 & M14 & M15 \\
\hline 
 $\rho_{c}$(g $cm^{-3}$ x 10$^{-11}$) & 1.16 & 17.54 & 3.23 & 18.04 & 4.81 & 16.40 & 56.91  \\

$T_{c}$(K) & 542 & 1017 & 521 & 1424 & 938 & 2140 & 5860   \\ 
 
 $R_{fragment}$(AU) & 17 & 6 & 10 & 7 & 11 & 9 & 8   \\
 
 $\dot M_{\rm final}(M_{\odot}$ yr$^{-1}$ x 10$^{-6}$) & 0.386 &  14.110  &  14.480  & 2.212 & 0.102 & 1.200  & 1.270 \\
 $M_{env}/M_{c} (\%)$ & 92.66 & 71.25 & 95.35 & 69.44 & 69.73 & 69.68 & 70.03 \\
\hline
\end{tabular}
\end{table*}

\section{Results and Discussion}

\begin{table*} \label{tbl-5}
\centering
 \caption{Summary of the sum of the perturbation amplitudes, the type of binary components, the binary separation $d$, the semi-major axis $a$, the eccentricity $e$, and the mass ratio $q$ for each model that evolved into a binary system. }
 \begin{tabular} {cccccccc}
\hline
\hline
 Model & ($A_{1}+A_{2}$) & Type of components & $d$ (AU) & $a$ (AU) & $e$ & $q$ \\
 \hline
M1 & 0.150 & VLMS - planemo & 144.00 & 121.70 & 0.454 & 0.150 \\
M1 & 0.150 & BD - planemo & 241.30 & 143.80 & 0.939 & 0.172 \\
M1 & 0.150 & VLMS - BD & 366.80 & 276.50 & 0.889 & 0.871 \\
M2 & 0.125 & BD - BD & 100.50 & 428.90 & 0.793 & 0.742 \\
M2 & 0.125 & BD - BD & 116.60 & 179.30 & 0.409 & 0.302 \\
M2 & 0.125 & BD - BD & 209.10 & 194.30 & 0.078 & 0.294 \\
M2 & 0.125 & BD - BD & 137.20 & 233.80 & 0.413 & 0.886 \\
M2 & 0.125 & planemo - planemo & 119.90 & 219.60 & 0.531 & 0.921 \\
M2 & 0.125 & BD- planemo & 172.30 & 259.90 & 0.371 & 0.158 \\
M2 & 0.125 & BD- planemo & 202.30 & 237.70 & 0.168 & 0.171 \\
M2 & 0.125 & BD - BD & 122.40 & 185.30 & 0.382 & 0.224 \\
M2 & 0.125 & BD - BD & 177.50 & 94.86 & 0.935 & 0.676 \\
M3 & 0.300 & VLMS - VLMS & 206.40 & 441.40 & 0.720 & 0.971 \\
M5 & 0.375 & VLMS - VLMS & 141.90 & 321.50 & 0.575 & 0.988 \\
M6 & 0.225 & VLMS - VLMS & 258.90 & 564.10 & 0.796 & 0.907 \\
M7 & 0.150 & BD - BD & 231.30 & 113.90 & 0.913 & 0.706 \\
M9 & 0.075 & BD - BD & 134.00 & 157.10 & 0.668 & 0.422 \\
M10 & 0.075 & BD - BD & 183.50 & 340.00 & 0.575 & 0.857 \\
M10 & 0.075 & VLMS - BD & 176.30 & 392.60 & 0.607 & 0.526 \\
M13 & 0.075 & BD - BD & 78.92 & 267.80 & 0.706 & 0.822 \\
M13 & 0.075 & BD - BD & 155.80 & 93.50 & 0.671 & 0.375 \\
M13 & 0.075 & BD - BD & 104.60 & 86.64 & 0.493 & 0.367 \\
M13 & 0.075 & BD - BD & 82.38 & 271.2 & 0.719 & 0.451 \\
M13 & 0.075 & BD - BD & 108.50 & 68.50 & 0.738 & 0.716 \\
M13 & 0.075 & BD - BD & 85.11 & 66.68 & 0.436 & 0.453 \\
M14 & 0.075 & VLMS - BD & 77.75 & 652.2 & 0.884 & 0.270 \\
M14 & 0.075 & VLMS - BD & 112.60 & 57.10 & 0.998 & 0.652 \\
M14 & 0.075 & VLMS - BD & 72.67 & 81.76 & 0.469 & 0.403 \\
\hline
\end{tabular}
\end{table*}

We now discuss our simulation results in detail. For visualization of the results of our simulations, we use the visualization tool SPLASH which is developed and made publicly available to the community by \cite{SPLASH}.

\subsection{Impact of density perturbation} 

Figures 1 \& 2 show face-on views of the evolution of the models M1 - M6 \& M7 - M12, respectively.
Details on the initial setup as well as the properties of the final state of the simulations are given in Tables 1 - 4. We use the Lane-Emden equation \citep[see e.g.][]{Liu1996} to estimate the central density and the central temperature of the most massive fragments at the end of the simulations. The central temperatures we compute here in Tables 3 and 4 are based on the assumption that the polytropic index $n$ = 1.5 (derived from a constant adiabatic index $\gamma = 5/3$) remains constant throughout the evolution of the polytrope (fragment) in our simulations. However, the adiabatic index remains a subject of change due to the different excitation levels attained by $H_{2}$ during the phases of gas collapse \citep{Masunaga2000}. There the authors have mentioned that the collapsing gas can be modeled by considering $\gamma = 5/3$ when the central density of the gas approaches 10$^{-12}$ g $cm^{-3}$ leading to the formation of the first core. As the gas collapses further and the central density becomes close to 10$^{-7}$ g $cm^{-3}$,  $\gamma$ takes a value of 7/5 due to the excitation of $H_{2}$ rotational levels. Beyond this central density of 10$^{-7}$ g $cm^{-3}$ the formation of the second core takes place and $\gamma$ then takes a value of 1.1 due to the collisional dissociation of $H_{2}$. We therefore state here as words of caution that our computed values for the central temperatures should be considered only with caution, given the constraint of our assumed constant value of $\gamma$ throughout the phases of gas collapse. The radius of the most massive fragment is approximated by considering the peak density attained by the collapsing gas while we estimate the mass through our clump finding algorithm as described in section 3. The results are reported in Tables 3 \& 4 where we adopt the following methodology to estimate the physical properties associated with the most massive fragment in each model. For a system with a polytropic equation of state, we have

\begin{equation} \label{POLYTROPICEOS}
P = K \rho^{\gamma} = K \rho^{1+(1/n)},
\end{equation}

where $K$, $\gamma$, and $n$ are defined as the polytropic constant, the adiabatic index and the polytropic index, respectively.

The mass of a general polytropic model is \citep{Liu1996}

\begin{equation} \label{approx}
M= 4\pi\rho_{c}R^{3}\left(-\frac{1}{z} \frac{d\omega}{dz}\right)_{z=z_{s}},
\end{equation}

where $\rho_{c}$, and $R$ are the central density and the radius of the polytrope. The dimensionless variable $z$ and the dimensionless gravitational potential $\omega$ are defined as

\begin{equation} \label{eta}
z=AR,
\end{equation}

\begin{equation} \label{eta}
A^{2}= \left(\frac{4\pi G}{(n+1)K}\right)\rho_{c}^{(n-1)/n},
\end{equation} 

and

\begin{equation} \label{eta}
\omega = \left(\frac{\rho}{\rho_{c}}\right)^{1/n} 
\end{equation}
where $\rho$ denotes the mean density.

From equation 13, \label{approx} we estimate the central density of the most massive fragment as
\begin{equation} \label{eta}
\rho_{c}= \frac{M}{4\pi R^{3}\left(-\frac{1}{z} \frac{d\omega}{dz}\right)_{z=z_{s}}}. 
\end{equation}

For a polytrope with index $n$ = 1.5 (i.e. for VLMS or BDs in our simulations), we have $z_{s}$ = 3.65375 and $z^{2} \omega^{\prime}_{z}$ = 2.71406 \citep[see e.g. Table 19.1 of][]{KippenhahnWeigert}.

Now for the central temperature of the polytrope, we have from the ideal gas law
\begin{equation} \label{eta}
P = \rho R_{g} T /\mu,
\end{equation}
where $P$, $\rho$, $R_{g}$, T, and $\mu$ are the pressure, the density, the gas constant, the temperature, and the mean molecular weight, respectively.

Using equation 12 we finally obtain

\begin{equation} \label{eta}
T_{c} = \frac{K \mu \rho_{c}^{(n+1)/n}}{\rho_{c} R_{g}}.
\end{equation}
  
Table 5 provides a summary of all the binary systems formed in our simulations along with their properties.
A first look at the results (see the summary in Tables 1 and 2) reveals that for the range of initial conditions explored in this paper, the 
gravitational collapse model seems to support the formation of systems which contain a combination of VLMS and BDs as a result of primary fragmentation within the molecular cores. 
Even for azimuthal perturbations with small overall amplitudes, the collapse always produces either multiple secondary fragments orbiting a single brown dwarf or a single brown dwarf
surrounded by spiral density waves propagating through a circumstellar disk which remains gravitationally stable until the end of the simulations has been reached.  
From an observational point of view, one of the most important findings of the high spatial resolution imaging surveys is that the mass ratio of both VLMS and BD binaries is strongly biased towards nearly equal mass binaries \citep{b32}. These surveys suggest that $\sim$ 68$\%$ of the binaries composed of BDs/VLMS detected through direct imaging have mass ratio q $\geq$ 0.8 \citep{b32}.
Comparing the models in Tables 1 and 2, we see that all models with initial perturbations which are sufficiently strong and nearly symmetric ($\frac{A_{2}}{A_{1}} \geq$ 0.5) and with rotational parameter $\beta{}$ $\sim$ 16 \%, indeed form binaries composed of VLMS and BDs. The properties of these binary systems at the end of the simulations are summarized in Table 5.

\subsection{Impact of rotation} 

Comparing models M2 \& M9, which differ only by the rate of rotation (the rotational parameter $\beta{}$ varies from 0.1628 to 0.2118) indicates that despite the asymmetry of the perturbation, stable BD binaries with extreme mass ratios will not be formed if the rotation of the cores is too slow. In Model M2, the fragments formed through direct fragmentation merge and form a single BD with a surrounding disk which becomes gravitationally unstable at the final stage of this model at t = 48.3 kyr. As a result, density waves are excited which further fragment into a cluster of BDs and planemos. This may potentially suggest that there may be a threshold value for the rotation rate, below which asymmetric binary systems will not be able to form as we explore further below.

Similarly, when comparing the results of models M1 \& M7, we find further indications for the impact of the rotation rate. These two models have identical perturbations ($A_{1}$ = 0.1, $A_{2}$ = 0.05) which favor the formation of the binaries. However, the slower rotation rate of model M1 leads to additional fragmentation in the vicinity of the binary, as can be seen in the first panel of Figure 1, which shows a planemo forming as a result of the fragmentation of density waves in the circumbinary disk at t = 47.9 kyr. 
On the other hand, when the rate of rotation is increased in model M7, no secondary fragments are formed and the final system is a BD binary with nearly equal masses at t = 58 kyr, as shown in the first panel of Figure 2. 
A further increase in the initial rate of rotation of the molecular cloud core in model M8 (see the second panel of Figure 2) yields a single BD with no hints of any
secondary fragmentation in its surrounding disk when the final stage of the model is reached at t = 68.5 kyr. 

A similar trend can be observed with regard to fragmentation into multiple systems. 
This is illustrated by the final outcome of models M4 \& M10, which both have a relatively weak density perturbation ($A_{1}$ = 0.05, $A_{2}$ = 0.025). 
In both cases, the low strength of the perturbation allows strong density waves to develop around a primary fragment.
Secondary fragmentation of the density waves into planemos is only possible in model M4 at t = 51 kyr. Whereas a single VLMS and multiple BDs are the only fragments which form at the end of model M10 at t = 74.7 kyr as is evident from the corresponding panels in Figures 1 and 2. We believe that the fragmentation in model M10 does not go as far as in model M4 most likely because of the difference in rotational support. 

Models M11 \& M12 further illustrate the impact of the rotation rate on the outcome of the fragmentation process. While model M12 with its relatively slow rotation rate is still able to form VLMS alongside BDs and planemos at t = 59.1 kyr, model M11 only forms a single BD without secondary fragmentation. However, despite the fact that this model runs for a longer time until t = 73.2 kyr, it appears that the BD is still not able to grow to a VLMS.
  
     \begin{figure*} \label{fig:5}
    \centering
    \includegraphics[angle=0,scale=0.45]{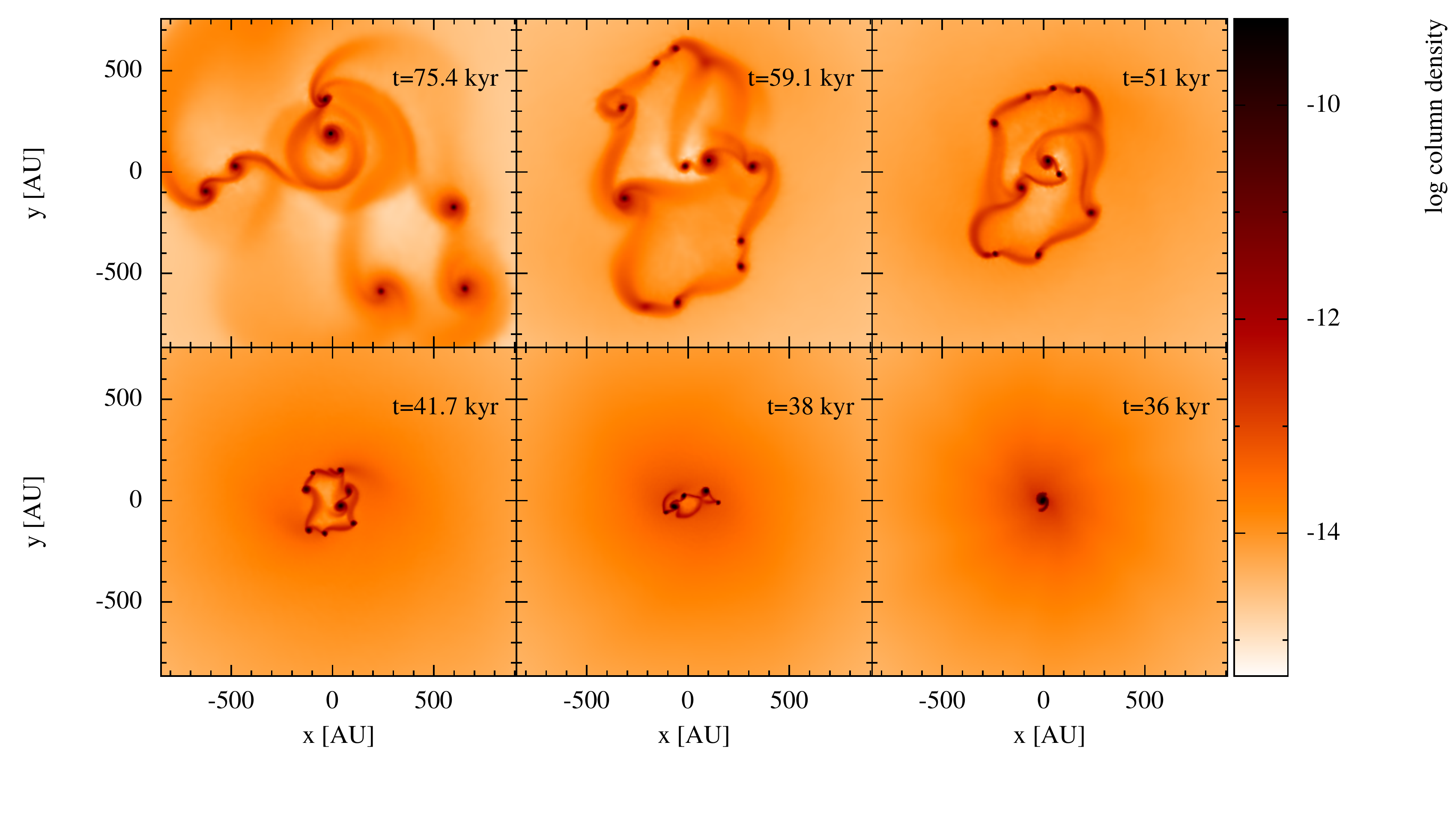}
    \caption{Simulation results for the sequence of models M10, M12, M4, M13, M14, \& M15 (panels from top left to bottom right) at the end of the computed evolution of each model. Each panel shows a face-on view of the column density integrated along the z-axis. The colour bar on the right shows log ($\Sigma{}$) in physical units of g cm$^{-2}$. The simulations are terminated when a total of $\sim$ 30 \% of the initial core mass is contained within the fragment(s) formed in each model. Each calculation was performed with 250025 SPH particles. }
  \end{figure*}

The embryo-ejection scenario allows for the ejection of single objects as well as a small fraction of close binaries. The ejection model finds that for the very survival of an ejected binary it must have a semi-major axis at least a factor of 3 below the orbital separation among the fragments within the parent cloud \citep{b43}. Numerical simulations have produced results which have led to claims related to binary formation via ejection \citep{b4,b45}. Their numerical findings are in fairly close agreement with observational data. The authors of the two papers mentioned above, respectively, concluded that the majority of circumstellar disks have radii less than 20 AU due to truncation in dynamical encounters which is consistent with observations of the Orion Trapezium cluster and implies that most stars and brown dwarfs do not form large planetary systems, and that the accretion of gas and the ensuing momentum transfer during the dynamical decay of triple systems is able to produce the observed distribution of close binary brown dwarfs, making the ejection model a viable option as a brown dwarf formation scenario.
Similarly, BD binaries from decaying triple systems with a semi-major axis distribution between 0.2 and 8 AU have been predicted \citep{b45}. The ejection model, however, has difficulties producing soft binaries with semi-major axes
exceeding 100 AU \citep{Goodwina, Goodwin2004,Goodwin7, b45}, whereas our models indeed produce some of those binary systems.

We therefore assume that the continued evolution of the binaries with wide separations (semi-major axis within the range 57 - 441 AU) which are formed in our models could lead to some of the observed systems in the regime of young soft VLMS and BD binaries such as {MASS J11011926-7732383AB\/} at a separation of 242 AU \citep{b23}, {ph 162225-240515\/} at a separation of 243 AU \citep{b55,b56}, {UScoCTIO 108\/} at a projected separation of 670 AU \citep{b57}, {FU Tau A and B\/} at a separation of 800 AU \citep{b20}, and {ENIS-J055146.0-443412.2\/} at a separation of over 200 AU \citep{b59}, respectively. Our findings can therefore shed some light on the possible initial conditions which are responsible for producing these rare wide VLMS-BD binary systems which account for no more than 1$\%$ to 2$\%$ of all systems. Our simulations suggest a formation scenario of wide VLMS and BD binaries which favours the collapsing molecular core scheme.

Models M5 and M6 show that stronger initial perturbation amplitudes 
lead to the formation of VLMS binary systems through direct fragmentation rather than BD binary systems.
These two models, together with model M3, also suggest that the overall level of the initial density
perturbation plays an important role in deciding whether additional secondary
fragmentation will be suppressed in the later stages of the evolution of collapsing
molecular cores. We suggest that if the total perturbation strength $A_{1}+A_{2}$ of the initial azimuthal density perturbation takes a value $< 0.22$, collapsing cores will more likely give birth to a BD binary system. A further decrease
in total perturbation strength enables secondary fragmentation within the disk(s) and consequently the emergence of planemos orbiting BDs and VLMS. 

As a note of caution, we remind the reader that the envelope surrounding the systems in models M1, M2, M3, M5, M6, M7, M8, M9 and M11 is not yet fully depleted at this point, so the evolution of these models is still subject to potentially further changes. Nevertheless, the current outcome shows quite a variety of possible systems, with systematic trends depending on the initial perturbations and the rotation rate of the cloud. While the method adopted here does not allow us to pursue the evolution much further as a result of very low timesteps, we note that the sink particle approach by \citet{b12} may allow further exploration of the evolution of such systems in future work \citep[see also][]{Riaz17}.

We also note that the BD system that emerged in model M7 has one component with a mass slightly less than the hydrogen burning limit and another with M $>$ 42.5 M$_{Jup}$. This binary system is consistent with the formation scheme proposed by \citet{b24}, which supposes that BD companions with masses above 42.5 M$_{Jup}$ are formed like stars through the fragmentation of molecular clouds similar to the formation of a stellar binary systems as we find 
here \citet{b24}. A list of the masses of the fragments in our simulations is provided in Table 2. 

Because our models only deal with the early formation stage of low mass binaries, they are relevant for the evolutionary stage called Common-Envelope evolution (CEE), which is the short-lived phase in the life of a binary star during which the two stars orbit inside a single, shared envelope \citep{b13}. 
We monitored the mass of the envelope M$_{env}$ and computed the ratio of this mass to the cloud mass M$_{cl}$ for each model. We define the envelope mass as the amount of matter in the cloud that does not belong to VLMS, BDs, and planemos. There are a few models such as M8, M9, \& M11 where high ratios ($>$ 90$\%$) of envelope mass to cloud mass are still present when these models have been terminated. The molecular gas residing in their envelopes might experience fragmentation if the simulations are evolved further. This matter could also be accreted by the central fragment, which is a BD in the models with high rotation rates, and eventually alter the fate of the system. We note, however, that other important effects like photoevaporation of the envelope may avoid the formation of a low-mass star rather than a substellar object, and so further investigation is needed to make sure whether such evolution scenarios are possible. 

Let us now focus on the set of models M4, M10, M12, M13, M1 and M15 in which we only vary the rotational parameter $\beta{}$ within the range (0.0045 $\leq$ $\beta{}$ $\leq$ 0.2658). The amplitudes of the initial density perturbations are fixed at ($A_{1}$ = 0.05, $A_{2}$ = 0.025).  
Figures 3 illustrates the comparison of the 6 models. This figure 3 shows face-on views of the column density evolution in the midplane of the models. Within the sequence of models M10, M12, M4, M13, M14, M15, the rotational parameter $\beta{}$ is progressively decreased within the range described above (see Table 1). 


 

 
Acccretion rates for substellar objects have been computed by \citet{Stamatellos08}. These authors suggest that their computed higher accretion rates ($\geqslant$ 10$^{-6}$ $M_{\odot}$ yr$^{-1}$) mainly occur during an early short-lived evolutionary phase ($\leqslant$ 15 kyr).
We also compute the mass accretion rates based on the difference of the accumulated mass of the most massive fragments divided by the time at two consecutive stages of evolution. On average this yields 4 x 10$^{-6}$ $M_{\odot}$ yr$^{-1}$ as the mass accretion rate for all the models explored here at their final stages of evolution (see the Tables 3 and 4). The mass accretion rate in most of the models is also found to be in close agreement with the magnitude of rate of accretion reported by \citet{Stamatellos08}.

  
The upper left panel in Figure 3 shows that the fragments in model M10, which is the model with the highest initial rotational parameter $\beta{}$ = 0.2658 are spread over the widest spatial space. Spiral density waves appear in circumstellar as well as in circumbinary disks. Since the disks touch each other, the density waves tend to interfere with each other. This model takes the longest time to terminate at t = 74.5 kyr. There are a total of 7 fragments among which there is a single VLMS and 6 BDs but no planemos. By the time a total of $\sim$ 30 \% of the initial core mass has become part of the fragments, the central density and the central temperature of the most massive fragment are $\rho_{c}$ = 17.54 x 10$^{-11}$ g $cm^{-3}$ and 1017 K, respectively. The core hosts two binary systems. One is of the type VLMS-BD with binary separation 176.30 AU and a semi-major axis of 392.60 AU. The eccentricity $e$ and the mass ratio $q$ of this binary system are 0.607 and 0.526, respectively. The other binary system is a BD-BD combination with a binary separation of 183.50 AU and a semi-major axis of 340.0 AU. The eccentricity and the mass ratio of the binary system are 0.575 and 0.857, respectively.

The second model M12 with initial rotational parameter $\beta{}$ = 0.2118 shows a relatively compact configuration of fragments compared to model M10. The spiral density waves associated with some of the fragments also have a tendency to develop into an interference pattern. In this model the system takes 59.1 kyr to terminate. There are 11 fragments including  a single VLMS and a single planemo. The other 9 fragments are BDs. At the final stage, the central density and the central temperature of the most massive fragment are $\rho_{c}$ = 18.04 x 10$^{-11}$ g $cm^{-3}$ and 1424 K, respectively. Despite the greater number of condensations, model M12 shows no fragments which pair up as binary systems. 

The third model is M4 with rotational parameter $\beta{}$ = 0.1628. Compared to models M10 \& M12, model M4 shows an even more compact configuration of fragments with interfering spiral density waves. The model terminates after 51 kyr. There are 10 fragments in the collapsing core including a single VLMS and a single planemo. The rest of the fragments are BDs. The central density and the Central temperature of the most massive fragment are $\rho_{c}$ = 10.19 x 10$^{-11}$ g $cm^{-3}$ and 1330 K, respectively. Again no binaries appear to have formed when the simulation has ended.

The fourth model is M13 with $\beta{}$ = 0.0849. The fragments in model M13 are confined within a radius of $\sim$ 300 AU. The spiral density waves within the circumstellar and circumbinary disks have a tendency to interfere. The model takes 41.7 kyr to terminate. At the end 8 BDs have appeared and the central density and central temperature of the most massive fragment are $\rho_{c}$ = 4.81 x 10$^{-11}$ g $cm^{-3}$ and 938 K, respectively. The system of BDs forms 6 gravitationally bound pairs which might evolve into binary systems. The separation of the most tightly bound BD-BD binary system is 78.92 AU and its semi major axis is 267.80 AU with eccentricity $e$ = 0.706 and mass ratio $q$ = 0.822. The separation of the most loosely bound BD-BD binary system is 155.80 AU and the semi major axis is 93.50 AU with $e$ = 0.671 and $q$ = 0.375, respectively. 

The fifth model is M14 with initial rotational parameter $\beta{}$ = 0.0321. In this model the fragments appear within a radius of $\sim$ 200 AU from the center of the core. The generation of spiral density waves is more suppressed than for the other faster rotating models. The model takes 38 kyr to terminate. The number of fragments is reduced to 4 including 2 VLMS and 2 BDs. The final central density and the central temperature of the most massive fragment are $\rho_{c}$ = 16.40 x 10$^{-11}$ g $cm^{-3}$ and 2140 K, respectively. The most tightly bound pair has a binary separation of 72.67 AU and and a semi-major axis of 81.76 AU with eccentricity $e$ = 0.469 and $q$ = 0.403, respectively. The most loosely bound binary system has a binary separation of 112.60 AU and a semi-major axis of 57.10 AU with $e$ = 0.998 and $q$ = 0.652. 

The last model in the series is M15 and has the smallest initial rotational parameter $\beta{}$ = 0.0045. Because of its weakest rotational support against gravity, model M15 produces only a single fragment that accretes material radially from the surrounding gas and becomes a VLMS. The central density and the central temperature of the massive fragment are $\rho_{c}$ = 56.91 x 10$^{-11}$ g $cm^{-3}$ and 5860 K, respectively. This system evolves over a period of 36 kyr, which is shortest time-span of the series of models. There are no strong spiral density waves around the single fragment.

Figure 4 illustrates the distribution of the masses of the fragments at the end of the simulations for models M4, M10, M12, M13, M14 and M15. We notice that only models M4 and M12, which have moderate values of the rotational parameter $\beta{}$, produce planemos. Remarkably, the models with rotation at both extremes of the explored range of the rotational parameter $\beta{}$ do not produce any planemo. 

\subsection{Binary properties}

Disk fragmentation provides a mechanism for the formation of soft VLMS-BD binary systems \citep{Goodwin07}. Recent observations have also revealed a small number of wide low-mass binaries in both field and young clusters \citep{b20}. \citet{Bate2005} have reported the formation of wide binary BD systems via the ejection mechanism in their simulations. The discovery of the binary system FU Tau with component masses of ~0.05 $M_{\odot}$ and ~0.015 $M_{\odot}$ and a projected separation of 800 AU in a unique (isolated) environment rather than in a stellar cluster or a stellar aggregate forces theorists to revisit their models for the origin of BDs \citep{b20}.
Our models M7 and M9 illustrate that the emergence of BD binary systems is indeed possible in isolation. However, we  caution that the systems in those two models are immature as $\sim$ 85 \% and $\sim$ 93 \% of the initial mass of the core is still part of the envelope. Nevertheless, the two systems which appeared in our simulations have semi-major axes $\sim$114 AU and $\sim$157 AU, respectively.  

In Figure 5 we present histograms of the distributions of the semi-major axis $a$, the eccentricity $e$, and the mass ratios $q$ associated with binary systems of the type VLMS-BD (yellow bars) and BD-BD (black bars), respectively. Within the set of models M4, M10, M12, M13, M14, M15, only M10, M13, \& M14 have produced binary systems and therefore we only include the binaries resulting from these models in the plots. In the first panel on the left we can see that the upper and lower limits for the semi-major axes of VLMS-BD and BD-BD binary systems are (57 AU, 652 AU) and (67 AU, 340 AU), respectively. Both types of binaries also show a tendency to form more tightly bound systems of $a$ $\sim$ 80 AU rather than loosely bound systems with $a$ $\sim$ 180 AU. 
The middle panel illustrates the trend related to the eccentricity associated with the two types of binaries. We find that the VLMS-BD systems cover a wide range of eccentricity (0.47 $\leq$ $e$ $\leq$ 1.0) whereas the BD-BD type binary systems exhibit a more narrow eccentricity range 0.47 $\leq$ $e$ $\leq$ 0.74. 
In the right panel of Figure 5, we finally compare the two types of binary systems with regard to their distribution of the mass ratios $q$. We find that both types of binary systems yield low and high mass ratio $q$. The  distribution of $q$ on average varies in the range 0.31 $\leq$ $q$ $\leq$ 0.74. This is relatively on the lower side if compared with the VLM primaries ($M_{primary}$ $<$ 0.1 $M_{\odot}$) produced by the radiation hydrodynamical calculation \citep{bate2012}.

        \begin{figure*} \label{fig:11}
    \centering
    \includegraphics[angle=0,scale=0.5]{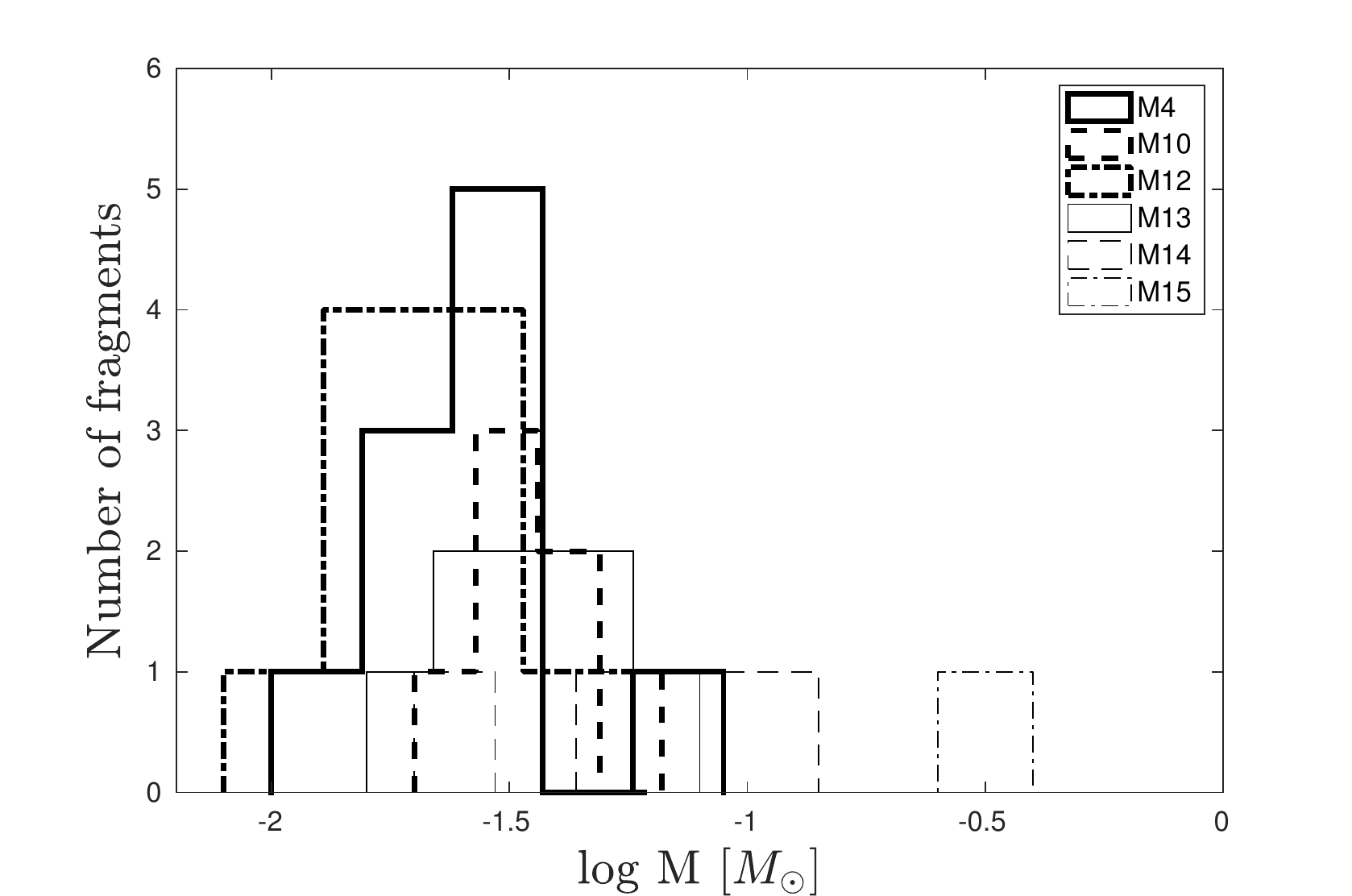}
    \caption{Distribution of the masses of the fragments which have been formed in models M4 (bold solid line), M10 (bold dashed line), M12 (bold dotted-dash line), M13 (thin solid line), M14 (thin dashed line) and M15 (thin dotted-dash line). The masses are given in units of solar mass $M_{\odot}$. The histograms are computed when a total of $\sim$ 30 \% of the initial mass of the core is contained by the fragment(s).}
  \end{figure*}

          \begin{figure*} \label{fig:12}
    \centering
    \includegraphics[angle=0,scale=0.5]{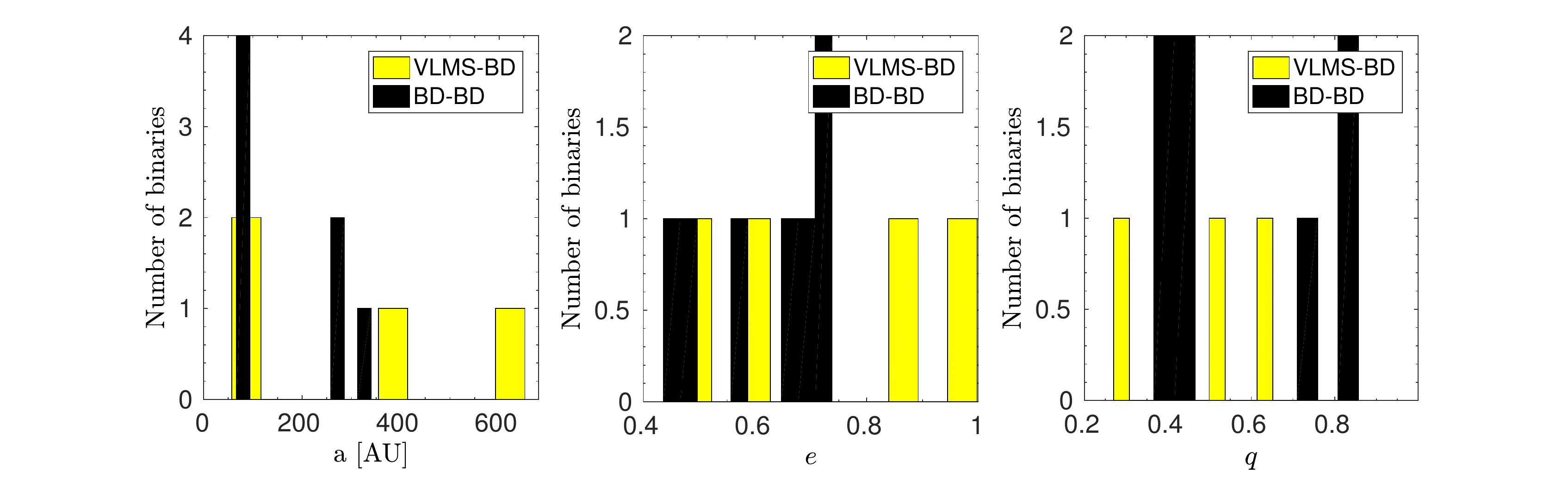}
    \caption{Distributions of the semi-major axis $a$ (left panel), eccentricity $e$ (middle panel), and mass ratio $q$ (right panel) of the binaries formed in models M10, M13, and M14. The yellow bars are for binary systems of the type VLMS-BD and the black bars are for binary systems which have been formed as BD pairs. The histograms are computed when a total of $\sim$ 30 \% of the initial mass of the core is contained by the fragment(s) formed in each model. 
    }
  \end{figure*}


\section{Limitations and restrictions}

The simulations presented in this work do not take into account radiative feedback effects on the collapsing gas. Stellar feedback can have a profound impact on the star forming gas  \citep{Krumholz2016, Myers2014, bate2012, Offner2009}. This has been shown, for example, in the comparative study performed by \cite{Bate2009b} in which smoothed particle hydrodynamics with radiative transfer \citep{WhitehouseBate2006} has been used. These simulations revealed a substantial decrease in the number of brown dwarfs formed when including radiative feedback effects. It is therefore necessary for our presented results to be compared with simulations including a radiative transfer scheme.

Based on the results of \cite{Offner2009}, we note in particular that the critical density for the transition towards adiabatic behavior shifts by about four orders of magnitude in density, implying a critical density of 10$^{-17}$ g $cm^{-3}$. Assuming the collapse in the previous isothermal regime follows approximately an isothermal sphere, the latter thus implies a change in the corresponding length scale by two orders of magnitude. This in turn implies that the basic processes modeled here are physical in their nature, but would usually occur at a different length scale. While current fragment radii in our models are thus of the order 10 AU, one may expect radii of 1000 AU when radiative feedback is taken into account. The latter has strong implications for the formation of binaries, and suggests that only wide binaries will be formed via gravitational collapse, thus complementing the stellar binaries at the long end where other models encounter difficulties. It is clear that the latter needs to be explored further at subsequent stages.

We intentionally do not explore the use of sink particles in our calculations as described already in section 4. Our method has the advantage that pressure forces between the protostars and the ambient gas are still taken into account, but comes at the price that the timestep decreases significantly during the formation of the clumps, and that the time evolution cannot be followed for as long. Nevertheless, we consider such complementary simulations as relevant to probe intermediate stages of the evolution and to examine the basic physics at work. Although we have shown that the current number of SPH particles in our simulations ensures the minimum resolvable density criterion, we still realize that there is a possibility that improved particle resolution and more extended time evolution \citep{Fryer2000, Klessen1998, bate1997} may refine the number of formed objects, BDs in particular, and their related properties. We also neglected the effects of magnetic fields, which may help to suppress fragmentation \citep[see e.g.][]{Peters2011, Banerjee2009, Price2009}. In addition, after the formation of an accretion disk, protostellar jets may be launched which can contribute to evacuate part of the gas from the environment and decrease the star formation efficiency \citep[see e.g.][]{Federrath2014, Price2012}.

\section{Conclusions and outlook}

The results reported in this paper strengthen the idea of fragmentation in collapsing molecular cores as the origin of VLMS and BD systems. The scheme successfully addresses the formation of wide VLMS and BD binaries with semi-major axis within the range 57 - 441 AU and covers a range of mass ratios close to the observed distribution \textit{q} $>$ 0.5. It seems likely that the physics involved in the formation of such wide binary systems (at least in the range explored here) will include core collapse and disk fragmentation. Planetary mass objects formed by gravitational fragmentation of circumstellar and circumbinary disks around VLMS or BDs also seem to be a frequent outcome of the cloud collapse scheme. 

Both the initial rate of rotation and the amplitude of the initial density perturbation are found to have a profound effect on the overall development of protobinary systems. Gravity,  hydrodynamics, and radiative feedback may be the primary ingredients for determining the statistical properties of low-mass star \cite{bate2014}. We in this work have followed the first two factors while neglecting the radiative feedback effect. Although, with sink particles based simulations \cite{bate2009} has reported negligible effects on the outcome from the changing the power spectrum of the initial velocity field at the molecular cloud level. We, by varying the initial density perturbation, however, mimic the variations in the turbulent structure of the gas at the molecular core level which has shown some effects on the properties of emerging fragmentation inside the collapsing gas core. Nevertheless, the effects on simulation outcome is shown more dependence on initial rate of rotation of the core than the initial density perturbation of the molecular gas inside the core. Asymmetric perturbations with mild rotation rates enable the formation of binary BD systems with extreme mass ratio. 
More symmetric perturbations lead to the commonly observed systems.

Systems with stronger perturbations can evolve into VLMS binaries. The multiplicity of the systems formed in our simulations strongly depends on the initial state of the cold molecular cloud cores. By constantly improving observational techniques such as gravitational microlensing, we may find further observational evidence to support and refine the theoretical work reported here. Further work should be focused on the comparison of binary population synthesis models to the observations, as well as including additional relevant physics such as the presence of magnetic fields and photoevaporation of the envelope in the CEE stage of binary formation.

Molecular cores evolving with relatively small rates of rotation can yield compact systems of VLMS and BDs. This may result in fragments influencing each other through viscous dynamical interactions associated with their disk structure such as tidal forces and disk collisions. These viscous interactions can generate retarding torques in the disk and eventually may cause orbital shrinkage. In this way slow rotating molecular cores may produce tightly bound binary systems.

Moderate values of rotational parameter $\beta{}$ produce planemos, while the models with rotational parameters at both extreme ends of the explored range do not produce planemos. 


The formation mechanism for very low mass stars or brown dwarfs described here may not only be relevant for very low mass stars in molecular clouds, but under a large range of physical conditions. For instance, some models assume that Pop. III stars were very massive, with masses $\sim$ 100~M$_\odot$ \citep{b60,b61}. However, it is conceivable that in such cases, fragmentation initially produces clumps of rather high masses, and three-body ejection events or other types of interaction may suppress accretion onto these protostars. Various investigations have indeed shown that fragmentation can occur (e.g. \citep{b62,b63,b64,b65}). Both the ejection of such clumps as well as potential merger events, \citep[see e.g.][]{b66}. Especially ejection may give rise to the formation of low-mass primordial stars, which could survive until the present day and be discovered in surveys looking for the most metal poor stars in the Universe.

As an important note of caution, we note that radiative feedback has been neglected in our models, but will play a role after the first protostars are forming. As demonstrated e.g. by \cite{Offner2009}, the latter effectively implies a stiffening of the equation of state, which will subsequently suppress fragmentation after the formation of the first protostars, and change the characteristic length scales where the transition to an approximately adiabatic evolution occurs by two orders of magnitude. As a result, we expect radiative feedback to favor the formation of wide-period binaries, and to strongly suppress the short-period ones. We therefore conclude that gravitational collapse may be important for wide-period binaries, where other mechanisms based on ejections encounter their difficulties \citep[see e.g.][]{Bate2009c, ThiesKroupa, Konopacky2007,b59, b23, Reipurth01}.

\section{Acknowledgements}
The first author RR acknowledges the high performance computing cluster at the Abdus Salam Centre for Physics, Pakistan and is thankful to Dr. Basmah Riaz for the personal visit and discussion that helped to set the theme of this paper. This research has also made use of the high performance computing clusters Geryon2 and Leftraru. The first author RR gratefully acknowledges support from the Department of Astronomy of the University of Concepcion, Chile. The second author wishes to thank Prof. Dr. R. Keppens and Prof. Dr. S. Poedts for providing access to the KUL supercomputing cluster Thinking while developing and testing the code that was used in this work. He also gratefully acknowledges the support of the KUL HPC team. The third author DRGS and RR thank for funding through the Concurso Proyectos Internacionales de Investigaci\'on, Convocatoria 2015" (project code PII20150171). DRGS further thanks for funding via Fondecyt regular (project code 1161247), via the Chilean BASAL Centro de Excelencia en Astrof\'isica yTecnolog\'ias Afines (CATA) grant PFB-06/2007. DRGS also thanks for funding through CONICYT Programa de Astrononom\'ia Fondo ALMA-CONICYT 2016 31160001, CONICYT Programa de Astrononom\'ia Fondo Quimal 2017 QUIMAL170001, as well as CONICYT PIA ACT172033.        


\label{lastpage}


\begin{thebibliography}{99}

\bibitem[\protect\citeauthoryear{Abel et al.}{2002}]{b60} Abel, Tom; Bryan, Greg L.; Norman, Michael L. 2002, Science, 295(5552), 93
\bibitem[\protect\citeauthoryear{Bacciotti et al.}{2011}]{b2} Bacciotti, F., Whelan, E. T., Alcal\'{a}, J. M., Nisini, B., Podio, L., Randich, S., Cupani, G. 2011, ApJl , 737(2), L26. 
\bibitem[\protect\citeauthoryear{Banerjee}{2009}]{Banerjee2009} Banerjee, R., Vazquez-Semadeni, E., Hennebelle, P., \& Klessen, R. S. 2009, MNRAS, 398(3), 1082

\bibitem[\protect\citeauthoryear{Basu and Vorobyov}{2012}]{basu12} Basu, S., \& Vorobyov, E. I. 2012, ApJ, 750(1), 30.

\bibitem[\protect\citeauthoryear{Bate}{2014}]{bate2014} Bate M. R. 2014, MNRAS, 442(1), 285

\bibitem[\protect\citeauthoryear{Bate}{2012}]{bate2012} Bate M. R. 2012, MNRAS, 419(4), 3115

\bibitem[\protect\citeauthoryear{Bate}{2009}]{bate2009} Bate M. R. 2009, MNRAS, 397(1), 232.

\bibitem[\protect\citeauthoryear{Bate}{2009}]{Bate2009b} Bate M. R. 2009, MNRAS, 392(4), 1363

\bibitem[\protect\citeauthoryear{Bate}{2009c}]{Bate2009c} Bate M. R. 2009, MNRAS, 392, 509

\bibitem[\protect\citeauthoryear{Bate et al.}{2003}]{b4} Bate M. R., Bonnell I. A., Bromm V. 2003, MNRAS, 339, 577.
\bibitem[\protect\citeauthoryear{Bate et al.}{2002}]{b5} Bate M. R., Bonnell I. A., Bromm V., 2002, MNRAS, 332, L65.
\bibitem[\protect\citeauthoryear{Bate and Bonnell}{2005}]{Bate2005} Bate, M. R., \& Bonnell, I. A.  2005, MNRAS, 356, 1201.

\bibitem[\protect\citeauthoryear{Bate and Burkert}{1997}]{bateburkert1997} Bate, M. R., \& Burkert, A.  1997, MNRAS, 288(4), 1060

\bibitem[\protect\citeauthoryear{Bate et al.}{1995}]{bate1997} Bate, M. R., Bonnell, I. A., \& Price, N. M. 1997, MNRAS, 277(2), 362


\bibitem[\protect\citeauthoryear{Baumgardt, H. and Klessen, R. S.}{2011}]{b66} Baumgardt, H. and Klessen, R. S. 2013, MNRAS, 413(3), 1810
\bibitem[Bejar (2008)]{b57} Bejar, V. J. S., Zapatero Osorio, M. R., Perez-Garrido, A., et al. 2008, ApJ, 673, L185
\bibitem[Billeres et al. (2005)]{b59} Billeres, M., Delfosse, X., Beuzit, J. L., Forveille, T., Marchal, L., \& Martín, E. L. 2005, AA, 440, L55

\bibitem[\protect\citeauthoryear{Bonnell}{1994}]{bonnell1994} Bonnell, I. A. 1994, MNRAS, 269(3), 837


 
\bibitem[\protect\citeauthoryear{Boss}{2007}]{b6} Boss, A. P. 2007, ApJl ,661(1), L73.
\bibitem[\protect\citeauthoryear{Bromm et al.}{2002}]{b61} Bromm, Volker; COPPI, Paolo S.; LARSON, Richard B. 2002, ApJ, 564(1), 23

\bibitem[\protect\citeauthoryear{Burkert and Bodenheimer}{1993}]{b7} Burkert, A., Bodenheimer, P. 1993, MNRAS, 264(4), 798.

\bibitem[\protect\citeauthoryear{Chabrier et al.}{2014}]{b9} Chabrier, G., Johansen, A., Janson, M., Rafikov, R. 2014, arXiv preprint arXiv:1401.7559.
\bibitem[\protect\citeauthoryear{Chabrier et al.}{2000}]{b10} Chabrier, G., Baraffe, I., Allard, F.,  Hauschildt, P. 2000, ApJ , 542(1), 464
\bibitem[\protect\citeauthoryear{Chiu et al.}{2006}]{b11} Chiu, K., Fan, X., Leggett, S. K., Golimowski, D. A., Zheng, W., Geballe, T. R., Brinkmann, J. 2006, ApJ , 131(5), 2722.

\bibitem[\protect\citeauthoryear{Clark et al.}{2011}]{b62} Clark, Paul C.; Glover, Simon C. O.; Smith, Rowan J.; Greif, Thomas H.; Klessen, Ralf S.; Bromm, Volker. 2011, Science, 331(6020), 1040

\bibitem[Close (2007)]{b56} Close, L. M., Zuckerman, B., Song, I., et al. 2007, ApJ, 660, 1492 

\bibitem[\protect\citeauthoryear{Duchene and Kraus}{2013}]{Duchene2013} Duchene, G., \& Kraus, A., ARAA, 51, 269.

\bibitem[\protect\citeauthoryear{Federrath et al.}{2014}]{Federrath2014} Federrath, C., Schrön, M., Banerjee, R., \& Klessen, R. S. 2014, ApJ, 790(2), 128


\bibitem[\protect\citeauthoryear{Fryer et al.}{2000}]{Fryer2000} Fryer, C. L., \& Heger, A. 2000, ApJ, 541(2), 1033


\bibitem[\protect\citeauthoryear{Goodwin et al.}{2004a}]{Goodwina} Goodwin, S. P., Whitworth, A. P., \& Ward-Thompson, D., 2004, A\&A, 414, 633
\bibitem[\protect\citeauthoryear{Goodwin et al.}{2004b}]{Goodwin2004} Goodwin, S. P., Whitworth, A. P., \& Ward-Thompson, D., 2004, A\&A, 423, 169
\bibitem[\protect\citeauthoryear{Goodwin and Kroupa}{2007}]{Goodwin7} Goodwin, S., Kroupa, P., Goodman, A., \& Burkert, A. 2007, Protostars and Planets V, ed. B. Reipurth, D. Jewitt, K. Keil (Tucson: University of Arizona Press), 133
\bibitem[\protect\citeauthoryear{Goodwin and Whitworth}{2007}]{Goodwin07} Goodwin, S. \& Whitworth, A. P. 2007, A\&A, 466, 943.
\bibitem[\protect\citeauthoryear{Greif et al.}{2012}]{b63} Greif, Thomas H., et al. 2012, MNRAS, 424(1), 399

\bibitem[\protect\citeauthoryear{Hayashi}{1966}]{Hayashi1966} Hayashi, C. 1966, Annu. Rev. AA, 4(1), 171

\bibitem[\protect\citeauthoryear{Hubber et al.}{2013}]{b12} Hubber, D. A., Walch, S., Whitworth, A. P. 2013, MNRAS, stt128.

\bibitem[\protect\citeauthoryear{Ivanova et al.}{2013}]{b13} Ivanova, N., Justham, S., Chen, X., De Marco, O., Fryer, C. L., Gaburov, E., Webbink, R. F. 2013, AAR, 21(1), 1.

\bibitem[Jayawardhana (2006)]{b55} Jayawardhana, R. and Ivanov, V. D. 2006, Science, 313, 1279
\bibitem[\protect\citeauthoryear{Jeffries}{2005}]{b14} Jeffries R. D., Maxted P. F. L. 2005, AN, 326, 944.

\bibitem[\protect\citeauthoryear{Joergens}{2010}]{b15} Joergens, V., Müller, A., Reffert, S. 2010, AA, 521, A24. 

\bibitem[\protect\citeauthoryear{Kippenhahn and Weigert}{1990}]{KippenhahnWeigert} Kippenhahn, R., Weigert, A., \& Weiss, A. 1990,  Stellar structure and evolution (Vol. 282). Berlin: Springer-Verlag

\bibitem[\protect\citeauthoryear{Kirkpatrick}{2005}]{b16} Kirkpatrick, J. D. 2005,  Annu. Rev. AA , 43, 195.

\bibitem[\protect\citeauthoryear{Klessen et al.}{1998}]{Klessen1998} Klessen, R. S., Burkert, A., \& Bate, M. R. 1998, ApJL, 501(2), L205


\bibitem[\protect\citeauthoryear{Konopacky et al.}{2007}]{Konopacky2007} Konopacky, Q. M., Ghez, A. M., Rice, E. L., \& Duchene, G. 2007, ApJ,663, 394

\bibitem[\protect\citeauthoryear{Kraus et al.}{2011}]{Kraus11} Kraus, A. L., Ireland, M. J., Martinache, F., \& Hillenbrand, L. A. 2011, ApJ, 731(1), 8


\bibitem[\protect\citeauthoryear{Krumholz et al.}{2016}]{Krumholz2016} Krumholz, M. R., Myers, A. T., Klein, R. I., \& McKee, C. F. 2016, 460(3), 3272

\bibitem[\protect\citeauthoryear{Larson}{1969}]{Larson1969} Larson, R. B. 1969,  MNRAS, 145(3), 271

\bibitem[\protect\citeauthoryear{Latif M. A. and Schleicher, D. R. G.}{2015}]{b65} Latif, M. A.; Schleicher, D. R. G. 2015, AA, 578, A118
\bibitem[\protect\citeauthoryear{Latif et al.}{2013}]{b64} Greif, Latif, M. A., et al. 2013, MNRAS, 433(2), 1607

\bibitem[\protect\citeauthoryear{Liu et al.}{2011}]{b17} Liu, M. C., Deacon, N. R., Magnier, E. A., Dupuy, T. J., Aller, K. M., Bowler, B. P., Wainscoat, R. J. 2011, ApJl , 740(2), L32.


\bibitem[\protect\citeauthoryear{Liu}{1996}]{Liu1996} Liu, F. K. 1996, MNRAS, 281(4), 1197

\bibitem[\protect\citeauthoryear{Liu}{2010}]{b18} Liu, M. C., Dupuy, T. J., \& Leggett, S. K. 2010, ApJ,722(1), 311.
\bibitem[\protect\citeauthoryear{Luhman}{2013}]{b19} Luhman, K. L. 2013, ApJl , 767(1), L1. 
\bibitem[\protect\citeauthoryear{Luhman et al.}{2009}]{b20} Luhman, K. L., Mamajek, E. E., Allen, P. R., Muench, A. A., Finkbeiner, D. P. 2009, ApJ, 691(2), 1265.
\bibitem[\protect\citeauthoryear{Luhman et al.}{2005}]{b21} Luhman, K. L., Lada, C. J., Hartmann, L., Muench, A. A., Megeath, S. T., Allen, L. E., Fazio, G. G. 2005, ApJl, 631(1), L69.
\bibitem[\protect\citeauthoryear{Luhman}{2004}]{b22} Luhman, K. L. 2004, ApJ, 617(2), 1216.  
\bibitem[\protect\citeauthoryear{Luhman}{2004}]{b23} Luhman, K. L. 2004, ApJ, 614, 398
\bibitem[\protect\citeauthoryear{Ma and Ge}{2014}]{b24} Ma, B., Ge, J. 2014, MNRAS , 439(3), 2781. 

\bibitem[\protect\citeauthoryear{Masunaga et al.}{2000}]{Masunaga2000} Masunaga, H., \& Inutsuka, S. I. 2000, ApJ, 531(1), 350 

\bibitem[\protect\citeauthoryear{Maxted and Jeffries}{2005}]{b25} Maxted, P. F. L., Jeffries, R. D. 2005, MNRAS letter , 362(1), L45.
\bibitem[\protect\citeauthoryear{Mohanty et al.}{2005}]{b27} Mohanty, S., Jayawardhana, R., Basri, G. 2005, ApJ , 626(1), 498.
\bibitem[\protect\citeauthoryear{Monin et al.}{2013}]{b28} Monin, J. L., Whelan, E., Lefloch, B., Dougados, C., de Oliveira, C. A. 2013, arXiv preprint arXiv:1301.4387.
\bibitem[\protect\citeauthoryear{Monin et al.}{2010}]{b29} Monin, J.-L., Guieu, S., Pinte, C., et al. 2010, AA , 515, A91

\bibitem[\protect\citeauthoryear{Myers et al.}{2014}]{Myers2014} Myers, A. T., Klein, R. I., Krumholz, M. R., \& McKee, C. F. 2014, MNRAS, 439(4), 3420

\bibitem[\protect\citeauthoryear{Offner et al.}{2009}]{Offner2009} Offner, S. S., Klein, R. I., McKee, C. F., \& Krumholz, M. R. 2009, ApJ , 703(1), 131

\bibitem[\protect\citeauthoryear{Peters et al.}{2011}]{Peters2011} Peters, T., Banerjee, R., Klessen, R. S., \& Mac Low, M. M. 2011, ApJ , 729(1), 72


\bibitem[\protect\citeauthoryear{Phan-Bao et al.}{2013}]{b30} Phan-Bao, N., Lee, C. F., Ho, P., Mart\'{\i}n, E., Tho, D. D. 2013, EPJ Web of Conferences (Vol. 47, p. 14001). EDP Sciences.
\bibitem[\protect\citeauthoryear{Phan-Bao et al.}{2008}]{b31} Phan-Bao, N., Riaz, B., Lee, C. F., Tang, Y. W., Ho, P. T., Mart\'{\i}n, E. L., Shang, H. 2008, ApJl , 689(2), L141.
\bibitem[\protect\citeauthoryear{Phan-Bao et al.}{2005}]{b32} Phan-Bao, N., Mart\'{\i}n, E. L., Reyl\'{e}, C., Forveille, T., Lim, J. 2005, arXiv preprint astro-ph/0506621. 
\bibitem[\protect\citeauthoryear{Price}{2012}]{Pricereview} Price, D. J., 2012, Journal of computational physics, 231(3),759.

\bibitem[\protect\citeauthoryear{Price et al.}{2012}]{Price2012} Price, D. J., Tricco, T. S., \& Bate, M. R. 2012, MNRASL, 423(1), L45


\bibitem[\protect\citeauthoryear{Price and Bate}{2009}]{Price2009} Price, D. J., \& Bate, M. R. 2009, MNRAS, 398(1), 33

\bibitem[\protect\citeauthoryear{Price and Monaghan}{2007}]{b33} Price, D. J., Monaghan, J. J. 2007, MNRAS , 374(4), 1347.
\bibitem[\protect\citeauthoryear{Price}{2007}]{SPLASH} Price D., 2007, PASA, 24, 159

\bibitem[\protect\citeauthoryear{Reipurth and Clarke}{2001}]{Reipurth01} Reipurth, B., \& Clarke, C. 2001, AJ, 122(1), 432.
\bibitem[\protect\citeauthoryear{Reipurth}{2000}]{b35} Reipurth, B., 2000, AJ, 120, 3177.
\bibitem[\protect\citeauthoryear{Riaz et al.}{2017}]{Riaz17} R. Riaz, S. Vanaverbeke, \& D.R.G. Schleicher, 2017, arXiv preprint astro-ph/1712.09646
\bibitem[\protect\citeauthoryear{Riaz et al.}{2014}]{b36} Riaz, R., Farooqui, S. Z., Vanaverbeke, S. 2014, MNRAS, 444(2), 1189.     
\bibitem[\protect\citeauthoryear{Riaz et al.}{2012}]{b37} Riaz, B., Lodieu, N., Goodwin, S., Stamatellos, D., Thompson, M. 2012, MNRAS, 420(3), 2497.
\bibitem[\protect\citeauthoryear{Ricci et al.}{2012}]{b38} Ricci, L., Testi, L., Natta, A., Scholz, A.,  de Gregorio-Monsalvo, I. 2012, ArXiv e-prints. 
\bibitem[\protect\citeauthoryear{Rigliaco et al.}{2011}]{b39} Rigliaco, E., Natta, A., Randich, S., et al. 2011, AA , 526, L6.
\bibitem[\protect\citeauthoryear{Scholz et al.}{2012}]{b40} Scholz, A., Jayawardhana, R., Muzic, K., Geers, V., Tamura, M., Tanaka, I. 2012, ApJ, 756(1), 24.
\bibitem[\protect\citeauthoryear{Scharf}{2009}]{b41} Scharf, C., Menou, K. 2009, ApJl, 693, 113.


\bibitem[\protect\citeauthoryear{Stamatellos and Whitworth}{2008}]{Stamatellos08} Stamatellos, D., \& Whitworth, A. P. 2008, MNRAS, 392(1), 413.

\bibitem[\protect\citeauthoryear{Stamatellos et al.}{2007}]{Stamatellos07} Stamatellos, D., Hubber, D. A., \& Whitworth, A. P.  2007, MNRAS, Letters, 382(1), L30.

\bibitem[\protect\citeauthoryear{Stumpf et al.}{2010}]{b42} Stumpf, M. B., Brandner, W., Joergens, V., Henning, T., Bouy, H., K\"{o}hler, R., Kasper, M. 2010, ApJ, 724(1), 1.
\bibitem[\protect\citeauthoryear{Thies}{2011}]{b43} Thies, I. 2011, Universitäts-und Landesbibliothek Bonn

\bibitem[\protect\citeauthoryear{Thies and Kroupa}{2008}]{ThiesKroupa} Thies, I., \& Kroupa, P. 2008, MNRAS, 390(3), 1200 



\bibitem[\protect\citeauthoryear{Toomre}{1964}]{b44} Toomre A., 1964, ApJ, 139, 1217
\bibitem[\protect\citeauthoryear{Umbreit}{2005}]{b45} Umbreit S., Burkert A., Henning T., Mikkola S., Spurzem R. 2005, ApJ, 623, 940.
\bibitem[\protect\citeauthoryear{Vanaverbeke et al.}{2009}]{b46} Vanaverbeke, S., Keppens, R., Poedts, S., Boffin, H. 2009, CPC , 180(7), 1164.
\bibitem[\protect\citeauthoryear{Veras}{2009}]{b47} Veras, D., Crepp, J. R., Ford, E. B. 2009, ApJ, 696, 1600.
\bibitem[\protect\citeauthoryear{Whelan et al.}{2009}]{b49} Whelan, E. T., Ray, T. P., Podio, L., Bacciotti, F., Randich, S. 2009, ApJ , 706(2), 1054.
\bibitem[\protect\citeauthoryear{Whelan et al.}{2005}]{b50} Whelan, E. T., Ray, T. P., Bacciotti, F., Natta, A., Testi, L., Randich, S. 2005, Nature, 435(7042), 652.
\bibitem[\protect\citeauthoryear{Whitworth et al.}{2007}]{b51} Whitworth, A., Bate, M. R., Nordlund, AA, Reipurth, B., Zinnecker, H. 2007, Univ.of Arizona Press, 459.
\bibitem[\protect\citeauthoryear{White and Hillenbrand}{2004}]{b53} White, R. J., Hillenbrand, L. A. 2004, ApJ, 616(2), 998.

\bibitem[\protect\citeauthoryear{White and Bate}{2006}]{WhitehouseBate2006} Whitehouse, S. C., \& Bate, M. R. 2006, MNRAS, 367(1), 32

\bibitem[\protect\citeauthoryear{Wright et al.}{2010}]{b54} Wright, E. L., Eisenhardt, P. R., Mainzer, A. K., Ressler, M. E., Cutri, R. M., Jarrett, T., Tsai, C. W. 2010, Aj , 140(6), 1868.






\end{thebibliography}
\end{document}